\documentstyle[festkoer,psfig]{vieweg}

\author{Johann Kroha and Peter W\"olfle}

\Kurzautor{J. Kroha and P. W\"olfle}

\title{Fermi and non-Fermi Liquid Behavior in Quantum Impurity Systems: 
Conserving Slave Boson Theory}

\Kurztitel{Fermi and non-Fermi Liquid Behavior}

\Adresse{Institut f\"ur Theorie der Kondensierten Materie, 
Universit\"at Karlsruhe\\ D-76128 Karlsruhe, Germany}

\begin{document}

\Titel

\begin{abstract}
We review a recently
developed method, based on an exact auxiliary boson representation,
to describe both Fermi liquid and non-Fermi liquid behavior 
in quantum impurity systems. Coherent spin and charge fluctuation processes
are taken into account in a self-consistent way and are shown to
include {\it all}
leading and subleading infrared singularities at any given order of 
the self-consistent loop expansion of the free energy.
As a consequence, for the SU(N)$\times$ SU(M) Anderson impurity models
the correct temperature dependence of the susceptibility is recovered 
over the entire temperature range, including
Fermi liquid or non-Fermi liquid behavior below the Kondo temperature $T_K$.
As a standard diagram technique the presented 
method has the potential to be generalized to correlated electron 
systems on a lattice. 
\end{abstract}

\section{Introduction}
Highly correlated electron systems are characterized by a strong repulsion 
between electrons on the same lattice site, effectively
restricting the dynamics to the Fock subspace of states without double 
occupancy of sites. The prototype model for such systems is the
Anderson impurity model, which consists of an electron in a localized 
level $\varepsilon _d < 0$ (called d-level in the following) 
with on-site repulsion $U$, hybridizing via a transition matrix 
element $V$ with one or several degenerate 
conduction electron bands or channels \cite{hewson.93}. 
Depending on the number of channels $M$,
the model exhibits the single- or the multi-channel
Kondo effect, where at temperatures $T$ below the Kondo temperature $T_K$ 
the local electron spin is 
screened ($M=1$) or overscreened ($M\geq 2$) by the conduction
electrons, leading to Fermi liquid (FL) 
or to non-Fermi liquid (NFL) behavior \cite{coxzawa.98} with characteristic 
low-temperature singularities, respectively.

As perhaps the simplest model to investigate the salient features of
correlations induced by short-range repulsion, the Anderson model plays a 
central role for the description of strongly correlated electron systems: 
In the limit of large spatial 
dimensions \cite{metzner.89} strongly correlated lattice systems reduce 
in general to a single Anderson impurity 
hybridizing with a continuum of conduction electron states whose properties 
are determined from a self-consistency condition imposed by the translational
invariance of the system \cite{kotliar.96}. Quantum impurity models have 
received further interest due to their relevance for mesoscopic systems like
single electron transistors or quantum point contacts \cite{schoen.97}.
Nonlinear conductance anomalies observed in the latter systems
\cite{ralph.94} have provided one of the strongest 
cases for the physical realization of the two-channel Kondo effect
generated by two-level systems with electron assisted tunneling. 

The above-mentioned systems call for the development of accurate and flexible 
theoretical methods, applicable to situations where exact solution methods
are not available. We here present 
a general, well-controlled auxiliary boson technique which correctly
describes the FL as well as the NFL case of the generalized SU(N)$\times$SU(M) 
Anderson impurity model. As a standard diagram technique 
it has the potential to be generalized for correlated lattice problems 
as well as for non-equilibrium situations in mesoscopic systems. 

In section 2 we describe several exact properties of the auxiliary
particle representation, while the conserving
slave boson theory is developed and evaluated in section 3.

\section{Exact Auxiliary Particle Representation}

\subsection{The SU(N)$\times$SU(M) Anderson Impurity Model}
The auxiliary or slave boson method \cite{barnes.76} is a powerful tool to 
implement the effective restriction to the sector of Fock space with
no double occupancy imposed by a large on-site repulsion $U$. 
The creation operator for an electron with spin $\sigma$ in 
the $d$-level is written in terms of
fermionic operators $f_{\sigma}$ and bosonic operators $b$ as 
$d^{\dag}_{\sigma} = f^{\dag}_{\sigma}b$.
This representation is exact, if 
the constraint that the total number operator of auxiliary fermions 
$f_{\sigma}$ and bosons $b$ is equal to unity is obeyed.
$f^{\dag}_{\sigma}$ and $b^{\dag}$ may be envisaged as creating the three
allowed states of the impurity: singly occupied with spin $\sigma$ or empty.

In view of the possibility of both FL and NFL behavior 
in quantum impurity systems mentioned in the introduction
it is useful to introduce $M$ degenerate channels for the
conduction electron operators $c^{\dag}_{\sigma\mu}$,
labeled $\mu = 1,2,\dots , M$, in such a way that in the limit of 
impurity occupation number $n_d \rightarrow 1$ (Kondo limit) the
$M$-channel Kondo model is recovered, i.e. the model obeys an
SU(M) channel symmetry. The slave bosons then form an SU(M) multiplet
$b_{\bar\mu}$ which transforms according to the
conjugate representation of SU(M), so that $\mu$ is a conserved quantum
number. Generalizing, in addition, to
arbitrary spin degeneracy $N$, $\sigma = 1,2,\dots , N$, one obtains the
SU(N)$\times$SU(M) Anderson impurity model in slave boson representation
\begin{equation}
H=\sum _{\vec k,\sigma ,\mu}\varepsilon _{\vec k}
c_{\vec k\mu\sigma}^{\dag}c_{\vec k\mu\sigma}+
E_d\sum _{\sigma} f_{\sigma}^{\dag}f_{\sigma}+
V\sum _{\vec k,\sigma ,\mu}(c_{\vec 
k\mu\sigma}^{\dag}b_{\bar\mu}^{\dag}f_{\sigma} +h.c.)\ ,
\label{sbhamilton}
\end{equation}
where the local operator constraint
$\hat Q \equiv \sum _{\sigma} f_{\sigma}^{\dag}f_{\sigma}+
              \sum _{\mu} b_{\bar\mu}^{\dag}b_{\bar\mu} = 1$
must be fulfilled at all times.

\subsection{Gauge Symmetry and Exact Projection onto the Physical Fock Space}
The system described by the auxiliary particle Hamiltonian 
(\ref{sbhamilton}) is invariant under simultaneous, local $U(1)$ gauge 
transformations, $f_{\sigma}\rightarrow f_{\sigma} {\rm e}^{i\phi 
(\tau )}$, $b_{\bar\mu}\rightarrow b_{\bar\mu} {\rm e}^{i\phi (\tau )}$, with 
$\phi (\tau )$ an arbitrary, time dependent phase. 
While the gauge symmetry guarantees the conservation of the local,
integer charge $Q$, it does not single out any particular $Q$, like $Q=1$.
In order to effect the projection onto the $Q=1$ sector of Fock space, 
one may use the following procedure \cite{abrikosov.65,coleman.84}:  
Consider first the grand-canonical ensemble with respect to $Q$ and
the associated chemical potential $-\lambda$.
The expectation value in the $Q=1$ subspace of any
physical operator $\hat A$ acting on the impurity states is then
obtained as 
\begin{equation}
\langle \hat A\rangle =
\lim _{\lambda \rightarrow \infty}
\frac {\frac{\partial }{\partial \zeta} \mbox{tr}
       \bigl[\hat A e^{-\beta (H+\lambda Q)} \bigr] _G} 
      {\frac{\partial }{\partial \zeta} \mbox{tr}
       \bigl[ e^{-\beta (H+\lambda Q)} \bigr] _G} =
\lim _{\lambda \rightarrow\infty}\frac{\langle \hat A\rangle _G}
{\langle Q \rangle _G}\ ,
\label{projection}  
\end{equation}
where the index $G$ denotes the grand canonical ensemble
and $\zeta$ is the fugacity $\zeta = {\rm e}^{-\beta\lambda}$.
In the second equality of Eq. (\ref{projection})  
we have used the fact that any physical operator
$\hat A$ acting on the impurity is composed of the impurity electron 
operators $d_{\sigma}$, $d^{\dag}_{\sigma}$, and thus annihilates 
the states in the $Q=0$ sector, $\hat A|Q=0\rangle =0$. 
It is obvious that the grand-canonical expectation value
involved in Eq. (\ref{projection}) may be factorized into auxiliary
particle propagators using Wick's theorem, thus allowing for
the application of standard diagrammatic techniques.

It is important to note that, in general, $\lambda$ plays the role of a 
time dependent gauge field. In Eq. (\ref{projection}) a time independent
gauge for $\lambda$ has been chosen. 
In this way, the projection is only performed at
one instant of time, explicitly exploiting the conservation of the local
charge $Q$. Thus, choosing the time independent gauge means that in the
subsequent development of the theory, the $Q$ conservation must be 
implemented exactly. This is achieved in a systematic way by means of
conserving approximations \cite{kadanoff.61}, i.e. by deriving all
self-energies and vertices by functional derivation from one common 
Luttinger-Ward functional $\Phi$ of the fully renormalized 
Green's functions, 
\begin{equation}
\Sigma_{b,f,c} = \delta \Phi \{G_b,G_f,G_c\} /\delta G_{b,f,c}.
\label{fderiv}
\end{equation}
This amounts to calculating all quantities of the theory
in a self-consistent way, but has the great advantage that gauge 
field fluctuations need not be considered.

\subsection{Infrared Threshold Behavior of Auxilary Propagators}
The projection onto the physical subspace, Eq. (\ref{projection}),
implies that the pseudo\-fermion and slave boson Green's 
functions $G_f$, $G_b$ are definied as the usual time-ordered, grand canonical 
expectation values of a pair of creation and annihilation operators, 
however evaluated in the limit $\lambda \rightarrow \infty$. 
It follows that the traces involved in  $G_f$, $G_b$ are taken purely over
the the $Q=0$ sector of Fock space, and thus the backward-in-time
contribution to the auxiliary particle propagators vanishes. 
Consequently, the auxiliary particle
propagators are formally identical to the core hole propagators
appearing in the well-known X-ray problem \cite{nozieres.69}, and the
long-time behavior of $G_f$ ($G_b$) is determined by the 
orthogonality catastrophe 
\cite{anderson.67} of the overlap of the Fermi sea without 
impurity ($Q=0$) and the fully interacting 
conduction electron sea in the presence  
of a pseudofermion (slave boson) ($Q=1$).  
It may be shown that the auxiliary particle
spectral functions have threshold behavior 
with vanishing spectral weight at $T=0$ for energies $\omega $ 
below a threshold $E_o$, and power law behavior above $E_o$,  
$A_{f,b}(\omega ) \propto \Theta (\omega - E_o ) \omega ^{-\alpha _{f,b}}$.

For the single-channel Anderson model, which is known to have a
FL ground state, the threshold exponents may be deduced
from an analysis in terms of scattering phase shifts, using the
Friedel sum rule,
since in the spin screened FL state the impurity acts as a pure
potential scatterer \cite{schotte.69,mengemuha.88,kroha.97,kroha.98}, 
\begin{eqnarray}
\alpha_f = \frac{2n_d - n_d^2}{N}\ ,\qquad
\alpha_b = 1-\frac{n_d^2}{N}\, \qquad\quad (N\geq 1, M=1)
\label{alpha_fb1}
\end{eqnarray}
These results have been confirmed by numerical renormalization 
group (NRG) calculations \cite{costi.94} and 
by use of the Bethe ansatz solution in connection with
boundary conformal field theory (CFT)
\cite{fujimoto.96}.
On the contrary, in the NFL case of the multi-channel Kondo model
the threshold exponents have 
been deduced by a CFT solution \cite{affleck.91} as
\begin{eqnarray}
\alpha_f = \frac{M}{M+N}\ ,\qquad
\alpha_b = \frac{N}{M+N}\, \qquad\quad (N\geq 2, M\geq N)
\label{alpha_fb2}
\end{eqnarray}
Since the dependence of $\alpha _f$, $\alpha _b$ on the impurity occupation 
number $n_d$ shown above originates from pure potential scattering,
it is characteristic for 
the FL case. The auxiliary particle threshold exponents are, therefore,
indicators for FL or NFL behavior in quantum impurity models of the
Anderson type.

\section{Conserving Slave Particle T-Matrix Approximation}
\subsection{Non-Crossing Approximation (NCA)}
The conserving formulation discussed in section 2.2 precludes mean field
approximations which break the $U(1)$ gauge symmetry, like slave boson mean 
field theory. Although the latter can in some cases successfully 
describe the low $T$ behavior of models with a FL ground state, 
it leads to a spurious phase transition at finite $T$ and, in particular,
fails to describe NFL systems.

\begin{figure}
\vspace*{-0cm}
\centerline{\psfig{figure=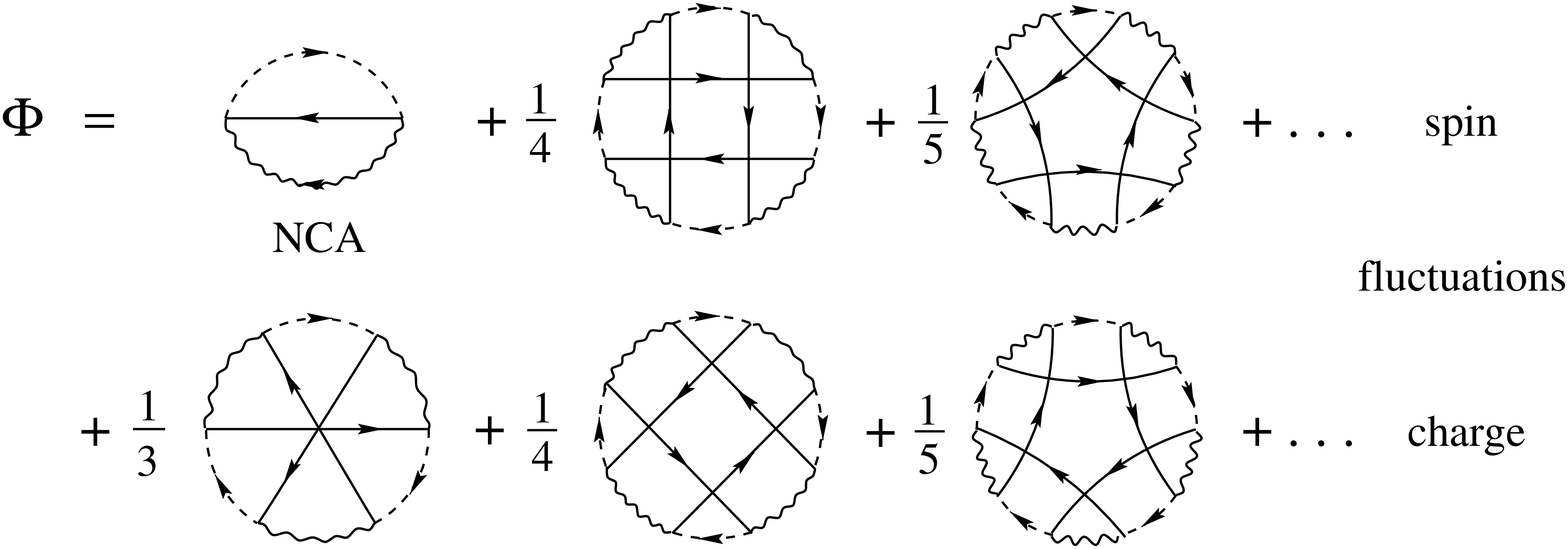,width=0.90\linewidth}}
\vspace*{0.0cm}
\caption{ 
Diagrammatic representation of the
Luttinger-Ward functional generating the CTMA. The terms with the conduction 
electron lines running clockwise (labelled ``spin fluctuations'') generate 
$T^{(cf)}$, while the terms with the conduction electron 
lines running counter-clockwise (labelled ``charge fluctuations'')
generate $T^{(cb)}$. The two-loop diagram is excluded,
because it is not a skeleton.}
\label{CTMA}
\end{figure}

Rather, the approximation should be generated from a Luttinger-Ward
functional $\Phi$. Using the hybridization $V$ as a small parameter,
one may generate successively more complex approximations. 
The lowest order conserving approximation generated in this way is
the Non-crossing Approximation (NCA) \cite{keiter.81,kuramoto.83}, 
defined by the first 
diagram in Fig. \ref{CTMA}, labeled ``NCA''. The NCA is successful in
describing Anderson type models at temperatures above and around the
Kondo temperature $T_K$, and even reproduces the threshold
exponents Eq. (\ref{alpha_fb2}) for the NFL case of the Anderson impurity
model. However, it fails to describe the FL regime at low temperatures.
This may be traced back to the failure to capture the spin-screened
Kondo singlet ground state of the model, since coherent 
spin flip scattering is not included in NCA, as seen below.

\subsection{Dominant Contributions at Low Energy}
\begin{figure}
\vspace*{-0cm}
\centerline{\psfig{figure=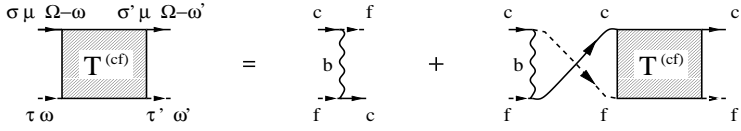,width=0.8\linewidth}}
\vspace*{0.0cm}
\caption{ 
Diagrammatic representation of the Bethe--Salpeter equation defining
the conduction electron--pseudofermion T-matrix $T^{(cf)}$. 
\label{tmat}}
\end{figure}
In order to eliminate the shortcomings of the NCA mentioned above, 
we may use as a guiding principle to look for contributions to the vertex 
functions which renormalize the auxiliary particle threshold exponents 
to their correct values, since this is a necessary condition for the 
description of FL and NFL behavior, as discussed in section 2.3. 
As shown by power counting arguments \cite{coxruck.93},
there are no corrections to the NCA exponents in any finite 
order of perturbation theory. Thus, any renormalization of the NCA exponents
must be due to singularities arising from an infinite resummation of terms. 
In general, the existence of collective excitations leads to a singular
behavior of the corresponding two--particle vertex function.  
In view of the tendency of Kondo systems to form a collective spin singlet
state, we expect a singularity in the spin singlet channel
of the pseudofermion--conduction electron vertex function.
It is then natural to perform a partial 
resummation of those contributions which, 
at each order in the 
hybridization $V$, contain the maximum number of spin flip processes. 
This amounts to calculating the conduction electron--pseudofermion 
vertex function in the ``ladder'' or T-matrix approximation, $T^{(cf)}$, 
where the irreducible vertex is given by $V^2G_b$.
The Bethe--Salpeter equation for $T^{(cf)}$ reads (Fig.~\ref{tmat}), 
\begin{eqnarray}
T^{(cf)\ \mu}_{\sigma\tau,\sigma '\tau '}
(i\omega _n, i\omega _n ', i\Omega _n ) =
&+&V^2G_{b\bar\mu}(i\omega _n + i\omega _n ' - i\Omega _n ) 
\delta _{\sigma\tau '}\delta _{\tau \sigma '}
\nonumber\\
&-&V^2T\sum _{\omega _n''}G_{b\bar\mu}(i\omega _n + i\omega _n '' - 
i\Omega _n ) \times 
\label{cftmateq}\\
&&\hspace*{-0.6cm}
G_{f\sigma}(i\omega _n'') \ G^0_{c\mu\tau}(i\Omega _n -i\omega _n '')\ 
T^{(cf)\ \mu}_{\tau \sigma,\sigma '\tau '}(i\omega _n '', i\omega _n ', 
i\Omega _n ), \nonumber
\end{eqnarray}\noindent
where $\sigma$, $\tau$, $\sigma '$, $\tau '$ represent spin indices 
and $\mu$ a channel index.
A similar integral equation holds for the charge fluctuation T-matrix
$T^{(cb)}$; it is obtained from $T^{(cf)}$ by interchanging
$f_{\sigma} \leftrightarrow b_{\mu}$ and $c_{\sigma\mu} \leftrightarrow
c^{\dag}_{\sigma\mu}$.
Inserting NCA Green's functions for the intermediate state
propagators of Eq. (\ref{cftmateq}), 
we find at low temperatures and in the Kondo regime 
$(n_d {\buildrel >\over\sim}  0.7)$ 
a pole of $T^{(cf)}$ in the singlet channel
as a function of the center--of--mass (COM) frequency $\Omega$, at a 
frequency which scales with the Kondo temperature, 
$\Omega = \Omega_{cf} \simeq - T_K$. 
Similarly, the corresponding $T$-matrix $T^{(cb)}$ in the conduction
electron--slave boson channel, evaluated within the analogous 
approximation, develops a pole at negative values of $\Omega$ in the 
empty orbital regime $(n_d {\buildrel <\over\sim} 0.3)$.  In the mixed valence 
regime ($n_d \simeq 0.5)$ the poles in both $T^{(cf)}$ 
and $T^{(cb)}$ coexist. The appearance of poles in the two--particle 
vertex functions $T^{(cf)}$ and $T^{(cb)}$, which signals the formation 
of collective states, may be expected to influence the behavior of 
the system in a major way.

\subsection{Self-consistent Formulation}
\begin{figure}
\vspace*{-0cm}
\hspace*{2cm}{\psfig{figure=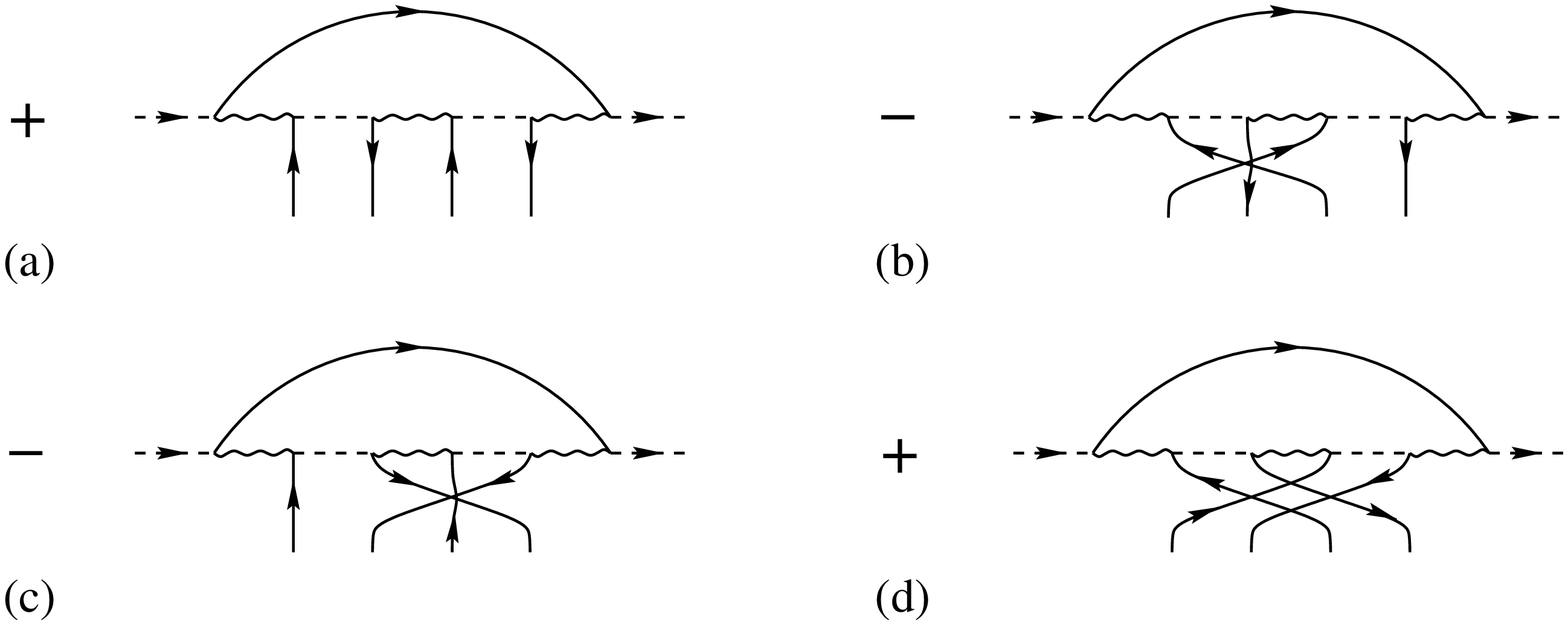,width=0.7\linewidth}}
\vspace*{0.0cm}
\caption{ 
Set of skeleton diagram parts {\em not}\/
contained in CTMA which cancel in the infrared limit
to leading and subleading order in the external frequency, 
$\omega \rightarrow 0$.   
\label{cancellation}}
\end{figure}
The approximation considered so far is
not yet consistent: The spectral weight produced by the poles
of $T^{(cf)}$ and $T^{(cb)}$ at negative frequencies $\Omega$
is strictly prohibited in the limit  $T\rightarrow 0$ 
by the threshold property of auxiliary particle vertex functions
(compare section 2.3). However, 
recall that a minimum requirement on the approximation used is the
conservation of gauge symmetry. This requirement is not met when the
integral kernel of the $T$-matrix equation is approximated by the NCA
result. Rather, the approximation should be generated
from the Luttinger-Ward functional shown in Fig.~\ref{CTMA}. 
It is defined as the infinite series of all vacuum skeleton diagrams 
which consist of a single ring of auxiliary particle propagators, where each
conduction electron line spans at most two hybridization vertices.
The first diagram of the infinite series of CTMA terms
corresponds to NCA. By functional 
differentiation with respect to the conduction electron Green's 
function and the pseudofermion or the slave boson propagator, 
respectively, the shown $\Phi$ functional generates the ladder 
approximations $T^{(cf)}$, $T^{(cb)}$ 
for the total conduction electron-pseudofermion vertex
function and for the total conduction electron-slave boson 
vertex function (Fig.~\ref{tmat}). The auxiliary particle
self-energies are obtained in the conserving scheme as the 
functional derivatives of $\Phi$ with respect to $G_f$ or
$G_b$, respectively (Eq.~(\ref{fderiv})), and are, in turn, 
nonlinear functionals of the full, renormalized auxiliary particle
propagators. This defines a set 
of self-consistency equations, which we term 
conserving T-matrix approximation (CTMA). 

The CTMA is justified on formal grounds by a cancellation theorem
for all diagrams not included in the generating 
functional $\Phi$ \cite{kroha.98}: As seen in Fig. \ref{CTMA}, $\Phi$
includes {\it all} contributions where a conduction electron line
spans up to two hybridization vertices; contributions not included
contain at least one conduction electron arch spanning four vertices.
These contributions cancel pairwise at arbitrary loop order 
to leading and subleading order in
the external frequency, as illustrated in Fig. \ref{cancellation}.

\begin{figure}
\vspace*{-0cm}
\centerline{\psfig{figure=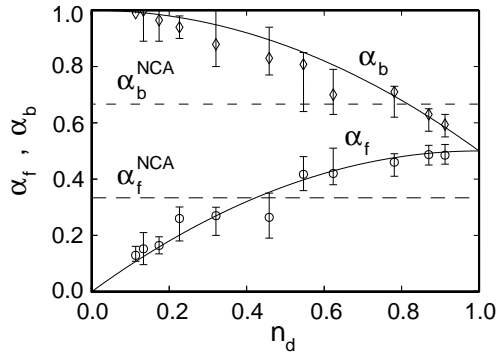,width=0.53\linewidth}}
\vspace*{0.0cm}
\caption{ 
The fermion and boson threshold exponents $\alpha _f$, $\alpha _b$
are shown for $N=2$, $M=1$ in dependence of the average impurity
occupation $n_d$. Solid lines: exact values, Eq. (\ref{alpha_fb1});
Symbols with error bars: CTMA; dashed lines: NCA. 
\label{exponents}}
\end{figure}
\subsection{Results}

We have solved the CTMA equations numerically for a wide range 
of impurity occupation numbers $n_d$ 
from the Kondo to the empty impurity regime     
both for the single-channel and for the two-channel Anderson
model down to temperatures of the order of at least $10^{-2} T_K$. 

The solution of the CTMA equations forces the T-matrices 
to have vanishing 
spectral weight at negative COM frequencies $\Omega$. Indeed, the
numerical evaluation shows that the poles of $T^{(cf)}$ and $T^{(cb)}$ 
are shifted to $\Omega = 0$ by self-consistency, where they merge 
with the continuous spectral weight present for $\Omega >0$, and thus 
renormalize the threshold exponents of the auxiliary spectral functions. 
For $N=2$, $M=1$ the threshold exponents $\alpha _f$, $\alpha _b$ 
extracted from the numerical solutions   
are shown in Fig.~\ref{exponents}. 
In the Kondo limit of the multi--channel case ($N\geq 2$, $M = 2,4$) 
the CTMA solutions are found not to alter the NCA values and reproduce the 
the correct threshold exponents, 
$\alpha _f = M/(M+N)$, $\alpha _b = N/(M+N)$.

\begin{figure}
\vspace*{-0cm}
\centerline{\psfig{figure=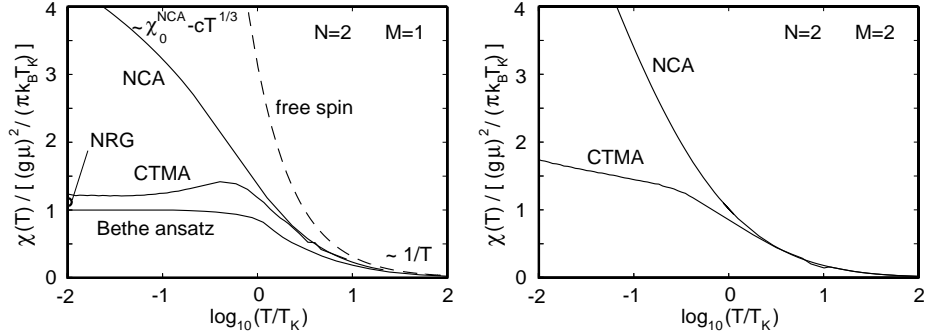,width=\linewidth}}
\vspace*{0.0cm}
\caption{
Static susceptibility of the single-channel ($N=2$, $M=1$) and the 
two-channel ($N=2$, $M=2$) Anderson impurity model in the Kondo regime 
($E_d=-0.8D$, $\Gamma = 0.1D$, Land\'e factor $g=2$). In the single-channel
case, the CTMA and NCA results 
are compared to the NRG result ($T=0$, same parameters) \cite{costi.help} 
and to the Bethe ansatz for the Kondo model \cite{andrei.83}. 
The CTMA susceptibility obeys scaling behavior in accordance
with the exact results (not shown).
\label{susc}} 
\end{figure}

The good agreement of the CTMA exponents with their exact values
over the complete range of $n_d$ for the
single--channel model and in the Kondo regime of the
multi--channel model may be taken as
evidence that the T-matrix approximation correctly 
describes both the FL and the non--FL
regimes of the SU(N)$\times$SU(M) Anderson model (N=2, M=1,2,4).
The static spin susceptibility $\chi$ of the single- and of the 
two-channel Anderson model in the Kondo regime calculated within CTMA
as the derivative of the magnetization 
$M = \frac{1}{2}g \mu _B \langle n_{f\uparrow} - n_{f\downarrow}\rangle$ 
with respect to a magnetic field $H$ is shown 
in Fig.~\ref{susc}. Good quantitative agreement with exact solutions is 
found for $N=2$, $M=1$ (FL). For $N=2$, $M=2$ (NFL) CTMA correctly reproduces
the exact \cite{andrei.84} logarithmic temperature dependence below 
the Kondo scale $T_K$. In contrast, the NCA solution recovers the 
logarithmic behavior only far below $T_K$. 

\section{Conclusion}
We have reviewed a novel technique
to describe correlated quantum impurity systems with strong onsite
repulsion, which is based on a conserving formulation of the 
auxiliary boson method. 
The conserving scheme allows to implement the
conservation of the local charge $Q$ without taking into account 
time dependent fluctuations of the gauge field $\lambda$. 
By including the leading infrared singular contributions
(spin flip and charge fluctuation processes),  
physical quantities, like the magnetic susceptibility, 
are correctly described both in the
Fermi and in the non--Fermi liquid regime,
over the complete temperature range, including the crossover to the 
correlated many--body state at the lowest temperatures.  
As a standard diagram technique this method has the potential to be 
applicable to problems of correlated systems on a lattice as well as
to mesoscopic systems out of equilibrium via the Keldysh technique.

We wish to thank S.~B\"ocker, 
T.A.~Costi, S.~Kirchner, A.~Rosch, A.~Ruckenstein and
Th.~Schauerte for stimulating discussions.
This work is supported by DFG through SFB 195 and by the
Hochlei\-stungsrechenzentrum Stuttgart.

\end{document}